\documentclass[aps,prd,reprint,amsmath,amssymb,superscriptaddress,floatfix]{revtex4}
\usepackage{graphicx}
\usepackage{amsfonts}
\usepackage{comment}
\usepackage[T1]{fontenc}
\usepackage{amsmath}
\usepackage{subfigure}
\usepackage{tikz,mathpazo}
\usepackage{amssymb}
\usepackage{rotating}
\usepackage{booktabs}
\usepackage{xcolor}
\usepackage{soul}
\usepackage{color}
\usepackage{slashed}
\usepackage{multirow}
\usepackage{makecell}
\usepackage{epsf}
\usepackage{ulem}
\usepackage{cancel}
\usepackage{color,bm}
\usepackage{bm}
\usepackage{mathtools}
\usepackage{diagbox}
\usepackage{array}

\usepackage[colorinlistoftodos]{todonotes}

\usepackage[colorlinks=true,citecolor=cyan,urlcolor=blue,bookmarks=true,bookmarks=true,bookmarksopen=true,bookmarksnumbered=true,bookmarksopenlevel=3]{hyperref}

\usepackage[compat=1.1.0]{tikz-feynman}

\definecolor{airforceblue}{rgb}{0.36, 0.54, 0.66}
\definecolor{steelblue}{rgb}{0.27, 0.51, 0.71}
\definecolor{amber}{rgb}{1.0, 0.49, 0.0}

\allowdisplaybreaks[4]

\begin{document}

\title{ The role of triangle singularity in the decay process $D^0 \to \pi^+ \pi^- f_0(980),\  f_0 \to \pi^+ \pi^-$ }

\author{Dazhuang He}
\author{Yiling Xie}
\affiliation{ Institute of Theoretical Physics, School of Physics, Dalian University of Technology, \\ No.2 Linggong Road, Dalian, Liaoning, 116024, P.R.China }
%\author{Xuan Luo}
%\affiliation{School of Physics and Optoelectronics Engineering, Anhui University, \\ Hefei, Anhui 230601, P.R.China}
\author{{Hao Sun}\footnote{haosun@dlut.edu.cn}}
\affiliation{ Institute of Theoretical Physics, School of Physics, Dalian University of Technology, \\ No.2 Linggong Road, Dalian, Liaoning, 116024, P.R.China }
%\date{\today}

\begin{abstract}

We study the process $D^0 \to \pi^+ \pi^- f_0(980),\ f_0 \to \pi^+ \pi^-$ by introducing the triangle mechanism, in which $f_0(980)$ is considered to be dynamically generated from the meson-meson interaction.
For the total contribution of this process,  the contribution of the triangular loop formed by $K^{*} \bar{K} K$ particles could generate a triangular singularity of about 1418 MeV.
%For the total contributions of this process, a triangle singularity around $1418$ MeV can only be generated from contribution of a triangle loop formed by $K^{*0} K^+ K^-$ particles.
We calculate the differential decay width of this process and show a narrow peak of about 980 MeV in the $\pi^+ \pi^-$ invariant mass distribution, which comes from $f_0$ decay. For the $M_{inv}(\pi f_0)$ invariant mass distribution, we obtain a finite peak at 1418 MeV, which is consistent with the triangle singularity.
%Finally, we obtain the decay branching ratio $\text{Br}(D^0 \to \pi^+ \pi^- f_0,f_0\to \pi^+ \pi^-)= 1.446 \times 10^{-4}$,
%which is expected to be compared with the experimental data.

\vspace{0.5cm}
\end{abstract}
\maketitle
\setcounter{footnote}{0}

\section{Introduction}
\label{I}

Triangle diagrams give the same good descriptions in the hadron physics, but of particular concern are those that lead to triangle singularities (TS) in amplitude \cite{Peierls:1961zz,Aitchison:1964zz,Bronzan:1964zz,Coleman:1965xm}. Triangle singularities were introduced by Landau \cite{Karplus:1958zz,Landau:1959fi}.
Nowadays, a large amount of peaks observed in high-energy experiments are considered to be caused by triangle singularities, especially the processes involving heavy quarks. With the development of experiments, triangle singularities play an increasingly important role in hadron physics.
%the role of  in hadron physics becomes more and more important.
The picture of the triangle mechanism can be summarized as follows: the initial particle $A$ decays into two internal particles 1 and 2 flying back-to-back, particle 2 decays into internal particle 3 and external particle B, the former moves in same direction as particle 1, and the two internal particles $1$ and $3$ re-scatter to form an external particle $C$. According to Coleman-Norton theorem \cite{Coleman:1965xm}, the generation of triangle singularity depends on whether the above process is a classical process and whether all three internal particles are simultaneously placed on-shell and collinear in the rest frame of the decay particle \cite{Karplus:1958zz}.
%In reality, the internal particles have finite widths and result in the transformation of triangle singularities to finite peaks which can be observed in experiments.
In reality, the internal particles, due to finite widths, can transform the triangle singularities to finite peaks which can be observed in experiments.

In Ref.\cite{Wu:2011yx}, the triangle singularity was introduced to explain $J/\psi \to \gamma \eta(1405/1475) \to \gamma \pi^0 f_0(980) \to \gamma 3\pi$ experiment data, and gave some good conclusions to explain the $f_0-a_0$ mixing and the relation between $\eta(1405/1475)$.
Soon after, some literature were published to analyze the important role of triangle singularity in the process $\eta(1405) \to f_0(980)\pi^0$ \cite{Aceti:2012dj,Wu:2012pg,Achasov:2015uua,Achasov:2018swa}.
Meanwhile, in order to further solve the issue of $f_0(980)-a_0(980)$ mixing and isospin breaking, the triangle mechanism in different processes that contain $f_0-a_0$ mixing has been researched \cite{Aceti:2015zva,Achasov:2016wll,Pavao:2017kcr,Aceti:2016yeb,Sakai:2017iqs,Liang:2017ijf,Mikhasenko:2015oxp}. In Ref.\cite{Dai:2018rra}, the production of $f_0(980)$ in semi-lepton decay process $\tau^- \to \nu_\tau \pi^- f_0(980)$ has been studied through a $K^* K^+ K^-$ triangle loop which produces a singularity located at $1418$ MeV in the $\pi f_0$ invariant mass distribution. Applying triangle mechanism to $f_1 (1285) \to \pi_0 a_0 (980)$ and $f_1 (1285) \to \pi_0 f_0 (980)$ processes and considering all three resonances as dynamically generated states, the authors in Ref.\cite{Aceti:2015zva} obtained the branching fractions which are consistent with the experiments. Especially in Ref.\cite{Pavao:2017kcr}, the authors found the two non-resonant peaks at $2850$ MeV in the invariant mass of $\pi D_{s0}$ pairs and around $3000$ MeV in the invariant mass of $\pi D_{s1}^+$ pairs, which are associated with the kinematical triangle singularities. By adjusting the values of  $K^*$ width, it shows the peak behavior of the real and imaginary parts of the loop function.

Recently, the authors in Ref.\cite{Feijoo:2021jtr} proposed the triangle singularity close to the $\bar{K} d$ threshold for the first time in the $p \Sigma^- \to K^- d$ and $ K^- d\to p \Sigma^-$ processes. In Ref.\cite{Yu:2021euw}, the processes $D^+_s \to a_0(980)\rho$ and $a_0(980)\omega$ including a $\pi^+ \pi^+ \eta$ loop have been researched.
Further, for the $\psi(2s) \to \pi^+ \pi^- K^+ K^-$ process, the authors in Ref.\cite{Huang:2021olv} introduced a moving triangle singularity in the range of 1.158  to 1.181 GeV.
In addition, a lot of other research has been done on the triangle mechanism \cite{Xie:2016lvs,Huang:2018wth,Roca:2017bvy,Debastiani:2017dlz,Samart:2017scf,Sakai:2017hpg,Xie:2017mbe,Bayar:2017svj,Liu:2015taa,Dai:2018hqb,Xie:2018gbi,Liang:2019jtr,Liu:2019dqc,Xie:2019iwz,Jing:2019cbw,Nakamura:2019nwd,Sakai:2020fjh,Ling:2021qzl,Molina:2020kyu,Wang:2016dtb,Hsiao:2019ait,Sakai:2020ucu,Debastiani:2018xoi,Oset:2018zgc,Ding:2020dio,Dai:2018zki, Huang:2020ptc,Huang:2018wgr}.

In this paper, following the procedure in Refs.\cite{Sakai:2017hpg,Pavao:2017kcr}, we study $D^0 \to \pi^+ \pi^- f_0$ and $D^0 \to \pi^+ \pi^- f_0,\ f_0\to \pi^+ \pi^-$ decay processes by introducing a triangle mechanism.
The amplitude for the production of $f_0(980)$ could be obtained by chiral unitary approach, which the $f_0$ is considered as dynamically generated from the meson-meson interaction \cite{Liang:2014tia,Oller:1997ti,Oller:1998hw}.
We consider the main contributions of $K^{*0}K^+ K^-$ triangle loop as well as the contributions that come from intermediate state $a_1^+(1260)$. %\rho \pi \pi$ and $\sigma \pi \pi$ triangle loops.
%{\color{red} Not only the traditional methods, Feynman parametrization and dispersion relation, can be used to calculate the loop integral to obtain the triangle amplitude, but also the authors in Ref.\cite{Bayar:2016ftu,Guo:2019twa} proposed a simple formula  for judging whether the triangular singularities exist, which derived by performing residue theorem for loop function, and analyzing the poles of the space integral. }
Although the traditional Feynman parameterization and dispersion relation methods could be used to calculate the loop integral to obtain the triangular amplitude, the authors in Ref.\cite{Bayar:2016ftu,Guo:2019twa} put forward a simple formula to judge whether the triangular singularity exists, such formula is derived by performing the residue theorem on the loop function, and the poles of the spatial integral are then analyzed.
%While considering the main contributions of $K^{*0}K^+ K^-$ triangle loop, we also consider the contributions come from $\rho \pi \pi$ and $\sigma \pi \pi$ triangle loop.
%{\color{red} In order to obtain the total amplitude of this process, the  vertices coupling strength of first different decay processes of $D^0$ have been calculated by fitting to corresponding experimental data.}
In order to obtain the total amplitude of this process, some coupling strength for different $D^0$ decay processes  need to be calculated by fitting to corresponding experimental data.
%Moreover, by calculating the total amplitude, the decay branching ratio of the underlying process has been obtained, which is possible to be compared with the existing experimental data in the certain error range.

The structure of this paper is as follows.
In Section \ref{II}, the detailed pictures of $D^0 \to \pi^+ \pi^- f_0(980),\  f_0 \to \pi^+ \pi^-$ decay process and interaction vertices including $D^0$ have been depicted.
We calculate the coupling strength of the $D^0 K^{*0} K^- \pi^+$ and $D^0 a_1^+ \pi^-$  vertices by fitting the corresponding experiments.
Then, we give the derivation details and formalism for the calculation of $D^0 \to \pi^+ \pi^- f_0(980),\ f_0 \to \pi^+ \pi^-$ decay that contains the triangle mechanism, where $f_0$ is considered as the dynamically generated states. Moreover, we give the expressions of the amplitudes for the corresponding Feynman diagrams.
%In Section \ref{III}, we show the numerical results, and the decay branching ratio has been obtained.
In Section \ref{III}, the numerical results for differential distribution of decay width as a function of the invariant masses have been shown.
Finally, a brief summary is given in Section \ref{IV}.

\section{Formalism}
\label{II}
\begin{figure}[h]
\begin{center}
\includegraphics[scale=0.8]{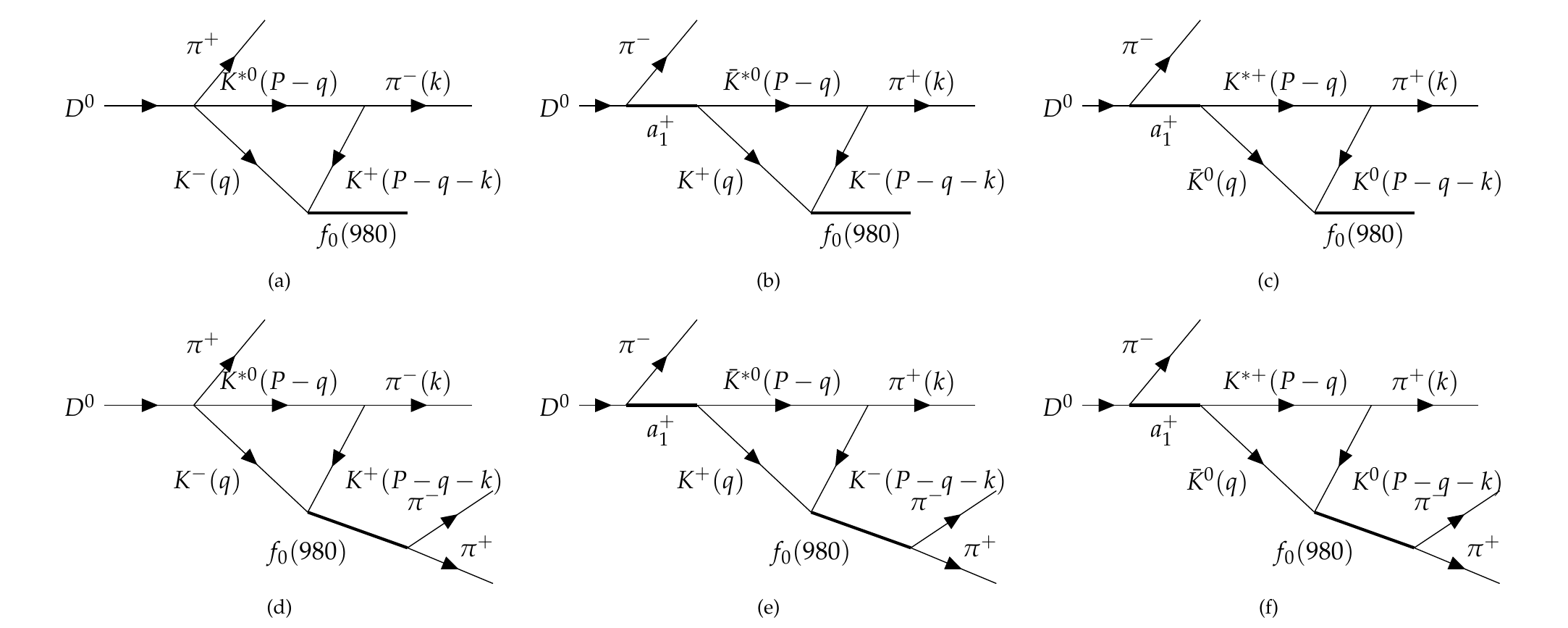}
\end{center}
\caption{
The total Feynman diagrams contributing to $D^0 \to \pi^+ \pi^- f_0(980)$ and $ D^0 \to \pi^+ \pi^- f_0(980), f_0(980)\to\pi^+ \pi^-$ decay processes involving intermediate state $a_1$ and $\bar{K}^{*} K \bar{K}$ triangle loop.}
\label{fig1}
\end{figure}

In this section, we show that a peak around 1.42 GeV in the $M_{inv}(\pi f_0)$ invariant mass distribution  will be produced in the $D^0 \to \pi^+ \pi^- f_0(980)$ and $D^0 \to \pi^+ \pi^- f_0(980), f_0 \to \pi^+ \pi^-$ decay processes by means of a  triangle singularity.
The total Feynman diagrams contributing to this process, which include the $\bar{K}^{*0}(K^{*0}) K^+ K^-$ and $K^{*+} \bar{K}^0 K^0$ triangle loops,  have been depicted in Fig.\ref{fig1}.
In order to determine the interaction strength of the first decay vertices of Fig.\ref{fig1}, such as  $D^0 \to \pi^+ K^{*0} K^- $ and $D^0 \to a_1^+ \pi^-$, the access to the corresponding decay width is essential.
From the PDG.\cite{Zyla:2020zbs}, for the first vertex  of $D^0 \to \pi K^* K$ process Figs.\ref{fig1}(a)(d), there is only the production channel of $\pi^\mp K^{*0} K^\pm $ pair.
Because the 3-rd component of isospin obeys the conservation law, only the Figs.\ref{fig1}(a)(d) will  contribute to the total amplitude for the first $D^0 \to \pi K^* K$ vertex.
%The PDG gives a branch ratio as Br$(D^0 \to a_1^+(1260), a_1 \to 2\pi^+ \pi^-) = (4.52 \pm 0.31) \times 10^{-3}$.
There is the similar situation for other subfigures of Fig.\ref{fig1}, but Ref.\cite{Wang:2015cis} gives the branching fractions of $ a_1^+ \pi^-$ and $a_1^- \pi^+$ productions, where one can see that the former is about two orders of magnitude larger than the latter.
Thus, for the vertices of $D^0 \to a_1^\pm \pi^\mp$ processes, we neglect the contribution from the $D^0 a_1^- \pi^+$ vertex.
Another important point is that the $a_1$ from the $D^0$ decay will be considered as dynamically generated in a two coupled channel problem with building blocks $\rho \pi$ and $\bar{K}^* K$ \cite{Wang:2015cis,Aceti:2016yeb}.
Thus, the $a_1$ could decay to $f_0 \pi$ by the $\bar{K}^* K \bar{K}$ and $\rho \pi \pi$ triangle loop  as Refs. \cite{Wang:2015cis,Aceti:2016yeb,Mikhasenko:2015oxp,Molina:2021awn}.
Because the width of $\rho$ meson is very broad, the contributions from the $\rho \pi \pi$ triangle loop are neglected \cite{Debastiani:2018xoi,Guo:2019twa}.
In the following, we will take Figs.\ref{fig1}(a)(d) as  example to perform the calculation and discussion.
From Fig.\ref{fig1}(a), we can see that the $D^0$  decays to $ \pi^+ K^{*0} K^-$ firstly, then the $K^{*0}$ decays to the $\pi^- K^+$.
Since the $K^+$ and $K^-$ move in the same direction, and the former is faster than the latter, the $K^+ K^-$ can be re-scattered to form $f_0(980)$.
Meanwhile, $f_0$ could be considered as the dynamically generated state of the $\pi^+ \pi^-$, $K^+ K^-$, $\pi^0 \pi^0$, $K^0 \bar{K}^0$ and $\eta \eta$ in $S$-wave within the chiral unitary approach \cite{Liang:2014tia,Oller:1997ti,Oller:1998hw}.

It is worthy to note that there is a triangle singularity for the $K^{*0} K^+ K^-$ triangle loop in Fig.\ref{fig1} when we take the value of the $f_0(980)$ mass just above the $K^+ K^-$ threshold, in other words, the mass $m_{f_0}$ must meet the condition for triangle singularity \cite{Bayar:2016ftu,Guo:2019twa}
\begin{equation}\label{condition}
\begin{aligned}
%m_{f_0}^2 \in \left[(2m_K)^2, \frac{Mm_K^2-m_{f_0}^2 m_K}{M-m_K}+M m_K  \right],
m_{f_0}^2 \in \left[(2m_K)^2, \frac{(m_K+m_{K^{*0}})(m_K m_{K^{*0}}  +m_K^2)-m_K m_{\pi} }{m_{K^{*0}}}  \right].
\end{aligned}
\end{equation}
%where the $M$ is the invariant mass of $K^{*0} K$ system.

In order to get the triangle singularity, it is straight to use the following condition
\begin{equation}\label{Eq:2-1}
\lim_{\epsilon \to 0}\left( q_{+}^{\text{on}} -q_-^{\text{a}} \right)=0,\quad \text{and} \quad q_{+}^{\text{on}}=\frac{\lambda^{\frac{1}{2}}(M_{inv}^2(\pi f_0),M_{K^+}^2,M_{K^{*0}}^2)}{2M_{inv}(\pi f_0)},
\end{equation}
where $q_{+}^{\text{on}}$ denotes the on shell three momentum of $K^{-}$ in the center of mass frame (COM) of $ K^{*0} K^-$, $M_{inv}(\pi f_0)$ is the invariant-mass of $ K^{*0} K^-$ system, and $\lambda(x, y, z)=x^2 +y^2 +z^2 -2xy-2yz -2xz$ is the K\"ahlen function. Meanwhile, $q_-^{\text{a}}$  is given by
\begin{equation}\label{Eq:2-2}
\begin{aligned}
q_-^{\text{a}} =\gamma(\nu E_{K^{+}}^*-p_{K^{+}}^*)-i \epsilon,
\end{aligned}
\end{equation}
with definitions
\begin{equation}\label{Eq:2-3}
\begin{aligned}
\nu =& \frac{k}{E_{f_0}}, \qquad \qquad \qquad \gamma=\frac{1}{\sqrt{1-\nu^2}}=\frac{E_{f_0}}{m_{f_0}},\\
E_{K^{+}}^*=&\frac{m_{f_0}^2+m_{K^{+}}^2-m_{K^-}^2}{2m_{f_0}},\ p_{K^{+}}^*=\frac{\lambda^{\frac{1}{2}}(m_{f_0}^2,M_{K^-}^2,M_{K^{+}}^2)}{2m_{f_0}},
\end{aligned}
\end{equation}
where $E_{K^{+}}^*$ and $p_{K^{+}}^*$ are the energy and momentum of the $K^{+}$ meson in the COM frame of the $K^{+}K^-$ system, $\nu$ and $\gamma$ are the velocity of the $K^+ K^-$ system and Lorentz boost factor, respectively.
When Eq.(\ref{Eq:2-1}) is established, on the one hand, it means that all particles in the triangle loop are on-shell, the $K^-$ and $f_0(980)$ move in the same direction in the rest frame of $f_0(980)$ and COM frame of $\pi f_0(980)$, respectively. Meantime, in order that the $K^+$ and $K^-$ in the triangle loop could re-scatter to form the $f_0(980)$ in a classical picture, it is required that the momentum of $K^-$ in the COM frame of $\pi  f_0$ is smaller than that of $K^+$ in the rest frame of the decay particle, which is derived from the $K^{*0} \to K^+ \pi^-$ decay.
On the other hand, in mathematics, Eq.(\ref{Eq:2-1}) represents that $q^a_-$ and $q^{\text{on}}_+$ are the singularities of triangle loop function in the upper and lower half of the complex-$q$ plane, respectively, and the integration path of loop function Eq.(\ref{Eq:b9}) is pinched between $q^a_-$ and $q^{\text{on}}_+$ in the same position of the real axis.
In addition,
\begin{equation}\label{Eq:2-4}
\begin{aligned}
E_{f_0}=\frac{M_{inv}^2(\pi f_0)+m_{f_0}^2-m_{\pi^+}^2}{2M_{inv}(\pi f_0)}, \qquad k=\frac{\lambda^{\frac{1}{2}}(M_{inv}^2(\pi f_0),m_{f_0}^2,m_{\pi^+}^2)}{2M_{inv}}.
\end{aligned}
\end{equation}
Now, we can obtain a triangle singularity around $1418$ MeV in the $M_{inv}(\pi f_0)$ invariant mass distribution by using Eq.(\ref{Eq:2-1}).
Further, when the complex mass of $K^{*0}$ with $m_{K^{*0}}^\prime=m_{K^{*0}}-i\Gamma_{K^{*0}}/2,\ \Gamma_{K^{*0}} = 50$ MeV is used in Eq.(\ref{Eq:2-1}), it leads to a complex triangle singularity near $1418-i29.7$ MeV.

\subsection{The process of $D^0 \to K^{*0} \pi K$}
\label{II-1}

\begin{figure}[h]
\begin{center}
\includegraphics[scale=0.9]{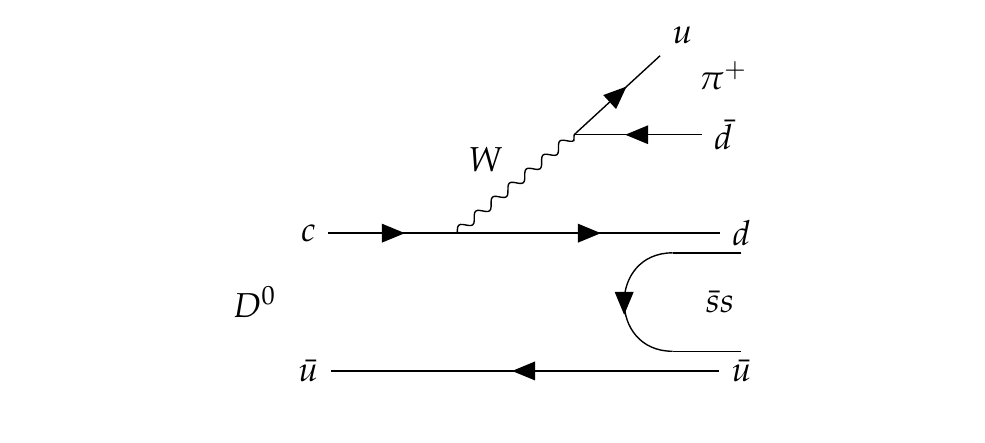}
\caption{
The $D^0 \to \pi^+ K^{*0} K^- $ decay process at the quark level.
%Because the 3-rd component of isospin obeys the conservation law, there is only $\pi^+ K^{*0} K^-$  final state.
}
\label{fig2}
\end{center}
\end{figure}

Before calculating the amplitudes of Fig.\ref{fig1}(a), the amplitudes of $D^0 \to \pi^{+} K^{*0} K^{-}$ decay processes need to be established to determine the vertex couplings strength.
The picture of the decay processes at the quark level is shown in Fig.\ref{fig2}, which only includes the $\pi^+ K^{*0} K^-$  final state.
Taking Fig.\ref{fig2} as an example, the first part is that the $\bar{d}u$ and $d\bar{u}$ quark pairs could be produced via an external emission of a $W$ boson,
where two vertices $cdW$ and $u\bar{d}W$ included by the weak decay are Cabibbo-suppressed and Cabibbo-favored, respectively.
In the next step, the $\bar{d}u$ quark-antiquark pair hadronizes and produces a $\pi^{+}$ meson.
At the same time, the remaining $d\bar{u}$ and $s\bar{s}$ quark-antiquark pairs, selected from the vacuum $\bar{q}q(\bar{u}u +\bar{d}d + \bar{s}s)$ state, combine to form the $K^{*0} K^-$ vector-pseudoscalar mesons pair.

In order to conserve angular momentum,  the coupling vertex of Fig.\ref{fig2} can be calculated via the $P$-wave interaction. %%%%we take the $D^0 \to K^{*0} \pi^+ K^-$ decay as an example,
Following the construction of vertex interactions in Refs.\cite{Sakai:2017iqs,Sakai:2017hpg}, we take
\begin{equation}\label{Eq:a1}
\begin{aligned}
-it_{D^0 \to K^{*0} \pi^+ K^- }=-iC \vec{\epsilon}_{K^{*0}} \cdot \vec{p}_{\pi^+},
\end{aligned}
\end{equation}
where $\vec{\epsilon}_{K^{*0}} $ and $ \vec{p}_{\pi^+}$ are the polarization vector of $K^{*0}$ and momentum of the $\pi^+$, respectively. % The concrete reason to take the form in Eq.(\ref{Eq:a1}) can be found in Refs.\cite{Sakai:2017hpg,Sakai:2017iqs}.
The $C$ presents the coupling strength of this vertex, the analytical expression and numerical results will be given below.
From Eq.(\ref{Eq:2-1}), we have obtained the triangle singularity around $1.42$ GeV, thus the momentum of $K^{*0}$ is approximately $135.66$ MeV in the $\pi^- f_0$ rest frame, which is smaller than the $K^{*0}$ mass $895.81$ MeV. For this reason, we can safely ignore the time-component $\epsilon^0$ of the $K^{*0}$ polarization vector, so the following form of polarization sum is taken
\begin{equation}\label{Eq:a2}
\sum_{\mu, \nu} \epsilon_{K^{*0}\mu}\epsilon_{K^{*0}\nu} \sim \sum_{i,j} \epsilon_{K^{*0}i}\epsilon_{K^{*0}j}=\delta_{ij}; \qquad \qquad  \mu=i,\ \mu=j;\qquad i,j=1,2,3.
\end{equation}

From the PDG \cite{Zyla:2020zbs}, the partial decay branching ratio are $\Gamma\left(D^0 \to K^{*0} K^{\mp} \pi^{\pm}, K^{*0} \to K^{\pm} \pi^{\mp}  \right)/\Gamma\left(K^+ K^- \pi^+ \pi^- \right)=11\pm 2\pm 1 \%$  and $\text{Br}\left(K^+ K^- \pi^+ \pi^- \right)=(2.45\pm 0.11)\times 10^{-3}$.
Considering the $\text{Br}\left(K^{*0} \to K\pi \right)\sim 100\%$, we can get $\text{Br}\left(D^0 \to K^{*0} K^{-}  \pi^{+} \right)=\frac{3}{2}\text{Br}\left(D^0 \to K^{*0} K^{-}  \pi^{+}, K^{*0} \to K^{+} \pi^{-} \right)$.
Meanwhile, the decay width of $D^0 \to \pi^+ K^{*0}  K^{-} $ process is given by
\begin{equation}\label{Eq:a3}
\begin{aligned}
\Gamma_{D^0 \to \pi^+ K^{*0} K^-} &=\int dM_{inv}( K^{*0} K^-)  \frac{1}{(2\pi)^3}\frac{|\vec{\tilde{p}}^\prime_{K^-} |  |\vec{\tilde{p}}_{\pi^+} |}{4m_{D^0}^2} \sum_{\text{pol}}|t_{D^0 \to \pi^+ K^{*0} K^-}|^2, \\
\end{aligned}
\end{equation}
where $M_{inv}( K^{*0} K^-)$ is the invariant mass of the $K^{*0} K^-$ system. The $\vec{\tilde{p}}^\prime_{K^-},\ \vec{\tilde{p}}_{\pi^+} $ are three momentum of the $K^-$ in the $K^{*0}K^-$ COM frame and that of the $\pi^+$ in the $D^0$ rest frame, respectively. They are given by
\begin{equation}\label{Eq:a4}
\begin{aligned}
|\vec{\tilde{p}}^\prime_{K^-} |&=\frac{\lambda^{\frac{1}{2}}\left(M_{inv}^2( K^{*0} K^-),\ M^2_{K^-},\ M^2_{K^{*0}}\right)}{2 M_{inv}( K^{*0} K^- )},\\
|\vec{\tilde{p}}_{\pi^+} |&=\frac{\lambda^{\frac{1}{2}}\left(M^2_{D^0},\ M_{inv}^2( K^{*0} K^-),\ M^2_{\pi^+} \right)}{2 M_{D^0}}.
\end{aligned}
\end{equation}
After squaring the amplitude $t_{D^0 \to \pi^+ K^{*0} K^- }$ and employing the polarization sum Eq.(\ref{Eq:a2}), we get
\begin{equation}\label{Eq:a5}
\sum _{\text{pol}} |t_{D^0 \to \pi^+ K^{*0} K^-}|^2=C^2 |\vec{\tilde{p}}_{\pi^+}^\prime|^2,
\end{equation}
where the $\vec{\tilde{p}}_{\pi^+}^\prime$ is the three momentum of the $\pi^+$ in the $K^{*0}K^-$ COM frame
\begin{equation}
|\vec{\tilde{p}}_{\pi^+}^\prime |=\frac{\lambda^{\frac{1}{2}}\left(M^2_{D^0},\ M_{inv}^2( K^{*0} K^-),\ M^2_{\pi^+} \right)}{2 M_{inv}( K^{*0} K^-)}.
\end{equation}
Finally, combining the Eq.(\ref{Eq:a3}) and Eq.(\ref{Eq:a5}), $C^2/\Gamma_{D^0}$ is given by
\begin{equation}\label{Eq:a6}
\begin{aligned}
\frac{C^2}{\Gamma_{D^0}}=\frac{\text{Br} \left( D^0 \to \pi^+ K^{*0} K^-\right) }{\int dM_{inv}( K^{*0} K^-)\frac{1}{(2\pi)^3}\frac{|\vec{\tilde{p}}^\prime_{K^-} |  |\vec{\tilde{p}}_{\pi^+} |}{4m_{D^0}^2} |\vec{\tilde{p}}_{\pi^+}^\prime|^2} \sim 5.69516 \times 10^{-7} \ \text{MeV}^{-1}.
\end{aligned}
\end{equation}

\subsection{The processes $D^0 \to \pi^+ \pi^- f_0(980)$ and $ D^0 \to \pi^+ \pi^- f_0(980), f_0(980)\to\pi^+ \pi^-$}
\label{II-2}

For the Feynman diagrams of Fig.\ref{fig1}(a) process, according to the Feynman rules, we have the following amplitude form
\begin{equation}\label{Eq:b1}
\begin{aligned}
-it_{D^0 \to \pi^+ \pi^- f_0}=i\sum_{\text{pol}} \int \frac{d^4 q}{(2\pi)^4} \frac{it_{D^0 \to \pi^+ K^{*0} K^-}}{q^2-m_{K}^2+i \epsilon} \frac{it_{K^{*0} K^+ \pi^-}}{(P-q)^2-m_{K^*}^2+i \epsilon}  \frac{it_{K^+ K^- f_0 } }{(P-q-k)^2-m_{K}^2+i \epsilon},
\end{aligned}
\end{equation}
in the COM frame of $\pi^- f_0$ where the $\pi^-$ comes from the $K^{*0}$ decay.
It can be seen from Eq.(\ref{Eq:b1}) that in order to obtain the full amplitude, the $t_{D^0 \to \pi^+ K^{*0} K^-},t_{K^{*0} K^+ \pi^-}$ and $t_{K^+ K^- f_0 }$ need to be calculated.
And, the $t_{D^0 \to \pi^+ K^{*0} K^-}$ has been obtained in Eqs.(\ref{Eq:a1}, \ref{Eq:a6}).

In order to calculate the $t_{K^{*0} K^+ \pi^-}$, we need to employ the chiral invariant Lagrangian with the local hidden symmetry\cite{Scherer:2002tk,Pich:1995bw}, which is given by
\begin{equation}\label{Eq:b2}
\mathcal{L}_{\text{VPP}}=-ig\langle V^\mu[P,\partial_\mu P]\rangle,
\end{equation}
where the brackets $\langle \cdots \rangle$ represents  SU(3) trace, and the coupling constant is given by $g=m_V/2f_\pi$ in the local hidden gauge, with $m_V=800$ MeV and $f_\pi=93$ MeV. The $P$ and $V^\mu$ stand for the pseudoscalar and vector mesons octet , respectively, which are given by
\begin{equation}\label{Eq:b3}
\begin{aligned}
&P=\begin{pmatrix}
\frac{\pi^0}{\sqrt{2}}+\frac{\eta_8}{\sqrt{6}} &\pi^+ &K^+\\
 \pi^-&-\frac{\pi^0}{\sqrt{2}}+\frac{\eta_8}{\sqrt{6}} &K^0\\
 K^-& \bar{K}^0& -\frac{2}{\sqrt{6}}\eta_8
\end{pmatrix}_, \qquad
&V=\begin{pmatrix}
	\frac{\rho^0}{\sqrt{2}}+\frac{\omega}{\sqrt{2}}& \rho^+& K^{*+}\\
	\rho^-&-\frac{\rho^0}{\sqrt{2}}+\frac{\omega}{\sqrt{2}} & K^{*0}\\
	K^{*-}& \bar{K}^{*0}& \phi
	\end{pmatrix}_.
\end{aligned}
\end{equation}
Then, the amplitude of $K^{*0}$ decay is written as
\begin{equation}\label{Eq:b4}
-it_{K^{*0} K^+ \pi^-}=-ig\vec{\epsilon}_{K^{*0}} \cdot (\vec{p}_{\pi^-}^\prime-\vec{p}_{K^+}^\prime).
\end{equation}
%The $K^{*0}$ in the triangle loop is  considered as on-shell and it's momentum is smaller than the mass leads that the zero component of polarization vector could be omitted, and the polarization sum Eq.(\ref{Eq:a5}) is used here.
Similarly to the calculations of Eq.(\ref{Eq:a5}), the zero component of polarization vector in Eq.\ref{Eq:b4} is omitted, and the polarization sum Eq.(\ref{Eq:a2}) is also used here.
The $\vec{p}_{\pi^-}^\prime$ and $\vec{p}_{K^+}^\prime$ are the momenta of the $\pi^-$ and $K^+$ in the  COM frame of $\pi f_0$, respectively, and the former is given by
\begin{equation}\label{Eq:b5}
\begin{aligned}
|\vec{p}_{\pi^-}^\prime|&=|\vec{k}|=\frac{\lambda^{\frac{1}{2}}\left(M^2_{inv}(\pi f_0),\ M_{\pi^-}^2,\ M^2_{f_0} \right)}{2 M_{inv}(\pi f_0)}.\\
\end{aligned}
\end{equation}
Finally, the $t_{K^+ K^- f_0 }$ need  to be provided, in which $f_0$ could be considered as the dynamically generated state. Thus the amplitude is simply written as
\begin{equation}\label{Eq:b6}
t_{K^+ K^- f_0 }=g_{K^+ K^- f_0 }.
\end{equation}

Now, by substituting Eqs.(\ref{Eq:a1},\ref{Eq:b4},\ref{Eq:b6}) into Eq.(\ref{Eq:b1}), we can get
\begin{equation}\label{Eq:b7}
\begin{aligned}
t_{D^0 \to \pi^+ \pi^- f_0}=i g_{K^+ K^- f_0 } g C \sum_{\text{pol}} \int \frac{d^4 q}{(2\pi)^4} \frac{\vec{\epsilon}_{K^*} \cdot \vec{p}_{\pi^+}^\prime }{q^2-m_{K}^2+i \epsilon}   \frac{\vec{\epsilon}_{K^*} \cdot (\vec{p}_{\pi^-}^\prime-\vec{p}_{K^+}^\prime) }{(P-q)^2-m_{K^*}^2+i \epsilon}  \frac{1}{(P-q-k)^2-m_{K}^2+i \epsilon} ,
\end{aligned}
\end{equation}
where $\vec{p}_{\pi^-}^\prime$ could be obtained from Eq.(\ref{Eq:b5}) and the $\vec{p}_{\pi^+}^\prime$ is the momentum of the $\pi^+$ in the COM frame of $\pi f_0$, which comes from the $D^0$ decay. We have
\begin{equation}
|\vec{p}_{\pi^+}^\prime|=\frac{\lambda^{\frac{1}{2}}\left(M^2_{D^0},\ M_{inv}^2( \pi f_0),\ M^2_{\pi^+} \right)}{2 M_{inv}( \pi f_0)}.
\end{equation}
The Eq.(\ref{Eq:a2}) could be employed to perform polarization sum of Eq.(\ref{Eq:b7}), which gives
\begin{equation}\label{Eq:b8}
\begin{aligned}
t_{D^0 \to \pi^+ \pi^- f_0}&=-i g_{K^+ K^- f_0 } g C  \int \frac{d^4 q}{(2\pi)^4} \frac{ 1 }{q^2-m_{K}^2+i \epsilon}   \frac{\vec{p}_{\pi^+}^\prime \cdot (2\vec{k}+\vec{q}) }{(P-q)^2-m_{K^*}^2+i \epsilon}  \frac{1}{(P-q-k)^2-m_{K}^2+i \epsilon}, \\
\end{aligned}
\end{equation}
where  $P=(M_{inv},0,0,0)$ and $\vec{p}_{\pi^-}^\prime-\vec{p}_{K^+}^\prime=\vec{k}-(-\vec{k}-\vec{q})=2\vec{k}+\vec{q}$.
We define the expression $t_T$ as
\begin{equation}\label{Eq:b9}
\begin{aligned}
t_T=i\int \frac{d^4 q}{(2\pi)^4} \ \vec{p}_{\pi^+}^\prime \cdot (2\vec{k}+\vec{q}) \frac{1 }{q^2-m_{K}^2+i \epsilon}   \frac{1 }{(P-q)^2-m_{K^*}^2+i \epsilon}  \frac{1}{(P-q-k)^2-m_{K}^2+i \epsilon}. \\
\end{aligned}
\end{equation}
For the $K^{*0}$ propagator in the $t_T$, we can ignore the part of negative energy as Refs.\cite{Aceti:2015zva,Bayar:2016ftu,Guo:2019twa}.  After performing analytically integration Eq.(\ref{Eq:b9}) in $dq^0$ by using residue theorem, we can use the following formula
\begin{equation}\label{Eq:b10}
\int d^3q \ \vec{q}_i\  f(\vec{q},\vec{k})=\vec{k}_i \int d^3 q \frac{\vec{q}\cdot \vec{k}}{|\vec{k}|^2} f(\vec{q},\vec{k}).
\end{equation}

Now, Eq.(\ref{Eq:b9}) reduces to
\begin{equation}\label{Eq:b11}
\begin{aligned}
t_T&= \vec{p}_{\pi^+}^\prime  \cdot \vec{k} \int \frac{d^3 q}{(2\pi)^3} \frac{1}{8\omega_{K^-}(\vec{q}) E_{K^*}(\vec{q}) E_{K^+}(\vec{k}+\vec{q})} \frac{1}{k^0- E_{K^*}(\vec{q})-E_{K^+}(\vec{k}+\vec{q})} \frac{1}{P^0+\omega_{K^-}(\vec{q}) +E_{K^+}(\vec{k}+\vec{q})-k^0}\\
&\times \frac{2P^0\omega_{K^-}(\vec{q})+2k^0 E_{K^+}(\vec{k}+\vec{q})-2\left[ \omega_{K^-}(\vec{q}) + E_{K^+}(\vec{k}+\vec{q})\right] \left[ \omega_{K^-}(\vec{q})+E_{K^*}(\vec{q})+E_{K^+}(\vec{k}+\vec{q})\right]}{\left[P^0- \omega_{K^-}(\vec{q}) - E_{K^+}(\vec{k}+\vec{q}) -k^0+i\epsilon \right] \left[P^0-\omega_{K^-}(\vec{q}) -E_{K^*}(\vec{q})+i\epsilon \right]}\left(2+  \frac{\vec{q}\cdot \vec{k}}{|\vec{k}|^2}\right),\\
&= \vec{p}_{\pi^+}^\prime  \cdot \vec{k} \times  \tilde{t}_T,
\end{aligned}
\end{equation}
where $\omega_{K^-}(\vec{q})=\sqrt{m_{K^-}+\vec{q}^2}, E_{K^{*0}}(\vec{q})= \sqrt{m_{K^{*0}}+\vec{q}^2}, E_{K^+}(\vec{q}+\vec{k})=\sqrt{m_{K^+}+(\vec{k}+\vec{q})^2}$ and $k^0=\sqrt{m_{\pi^-}^2+\vec{k}^2}$ are the energy of $K^-, K^{*0},K^+$ and $\pi^-$  in the COM frame of $\pi f_0$ system, respectively.
While $\ P^0=M_{inv}(\pi f_0)$ is the invariant mass of the $\pi f_0$ system.
The $|\vec{q}|$ integral in the Eq.(\ref{Eq:b11}) is regulated by the cutoff scheme, $\theta\left(q_{\text{max}}-|q^*| \right)$, where $|q^*|$ is the momentum of $K^-$ in the $f_0$ rest frame. The $q_{\text{max}}=600$ MeV in the $f_0$ rest frame was obtained in the chiral unitary approach by fitting experimental data. In the following calculation, we need to add the width for $K^{*0}$ with the replacement of $E_{K^{*0}}$ by $E_{K^{*0}}-i\Gamma_{K^{*0}}/2$ in the denominator of Eq.(\ref{Eq:b11}).

Finally, for the three body decay process as depicted in Fig.\ref{fig1}(a), the mass distribution of differential width could be written as
\begin{equation}\label{Gamma-1}
\begin{aligned}
\frac{d \Gamma^\prime}{d M_{inv}(\pi f_0)}&=\frac{1}{3}\frac{1}{(2\pi)^3}\frac{ C^2g_{K^+ K^- f_0 }^2 g^2}{4m_{D^0}^2} |\vec{p}_{\pi^+}||\vec{k}| |t_T|^2,\\
%&=\frac{1}{(2\pi)^3}\frac{1}{4m_{D^0}^2} |\vec{p}_{\pi^+}||\vec{k}| |\vec{p}_{\pi^+}^\prime|^2 |\vec{k}|^2 \frac{C^2}{3} g_{K^+ K^- f_0 }^2 g^2 |\tilde{t}_T|%^2,
\end{aligned}
\end{equation}
where the $1/3$ in Eq.(25) originate from the integrand of phase-space angle  in $\vec{p}_{\pi^+}^\prime  \cdot \vec{k}=|\vec{p}_{\pi^+}^\prime| | \vec{k}|\cos \theta$, and the $\vec{p}_{\pi^+}$ is the momentum of the $\pi^+$ in the $D^0$ rest frame
\begin{equation}\label{Eq:b12}
|\vec{p}_{\pi^+}|=\frac{\lambda^{\frac{1}{2}}\left(m_{D^0}^2,M_{\pi^+}^2,M_{inv}^2(\pi^- f_0) \right) }{  2m_{D^0}}.
\end{equation}

At the same time, the other Feynman diagrams contributions  Figs.\ref{fig1}(b)(c) were considered to calculate the differential decay width. Here we give the amplitude expressions ($t$) for the Figs.\ref{fig1}(a)(b)(c) directly
%\begin{equation}\label{M}
%\begin{aligned}
%t^{(b)(c)} & =  C^{\prime} g g_{ K^+K^- f_0} g_{ \bar{K}^{*}K a_1} \frac{1}{M_{inv}^2(\pi f_0)-m_{a_1}^2+im_{a_1}\Gamma_{a_1}}  \cdot t_{T} , \\
%\end{aligned}
%\end{equation}
\begin{equation}\label{M}
\begin{aligned}
t^{a} & =  -  C (C_2^a\ g)  \ \vec{p}_{\pi^+}^\prime  \cdot \vec{k} \times  \tilde{t}_T(m_{K^{*0}},m_{K^-}, m_{K^+}) \ (C_3^a\ g_{ \bar{K}K f_0}), \\
t^{b} & = - C^\prime \frac{1}{M_{inv}^2(\pi f_0)-m_{a_1}^2+im_{a_1}\Gamma_{a_1}} (C_1^b \ g_{ \bar{K}^{*}K a_1} ) (C_2^b\ g) \  \vec{p}_{\pi^-}^\prime  \cdot \vec{k} \times  \tilde{t}_T(m_{\bar{K}^{*0}},m_{K^+}, m_{K^-})\  (C_3^b\ g_{ \bar{K}K f_0}), \\
t^{c} & = - C^\prime \frac{1}{M_{inv}^2(\pi f_0)-m_{a_1}^2+im_{a_1}\Gamma_{a_1}} (C_1^c\ g_{ \bar{K}^{*}K a_1} ) (C_2^c\ g) \  \vec{p}_{\pi^-}^\prime  \cdot \vec{k} \times  \tilde{t}_T(m_{K^{*+}},m_{\bar{K}^0}, m_{K^0})\  (C_3^c\ g_{ \bar{K}K f_0}). \\
\end{aligned}
\end{equation}
From Refs.\cite{Aceti:2016yeb,Wu:2012pg,Wu:2011yx}, the values of the factors $C_i^j$ have been shown in Table.\ref{tabcij}.
\begin{table}[!htb]
\begin{center}
\renewcommand\arraystretch{2}
\begin{tabular}{| p{2cm}<{\centering} | p{1.5cm}<{\centering} |p{1.5cm}<{\centering}|p{1.5cm}<{\centering}|}
\hline
Diagram & $C_1 $ &  $C_2 $ &  $C_3 $ \\
\hline
Fig.\ref{fig1}(a) & \diagbox{}   &  1  &1 \\
\hline
Fig.\ref{fig1}(b) & $\frac{1}{\sqrt{2}}$ & 1 &1\\
\hline
Fig.\ref{fig1}(c) & -$\frac{1}{\sqrt{2}}$ & -1 &-1\\
\hline
\end{tabular}
\end{center}
\caption{ Coefficients entering the evaluation of amplitudes in Eqs.(\ref{M},\ref{Mdef}). The $C_2$ have a minus sign compared with Ref.\cite{Aceti:2016yeb}, this is because that we taken the  $\vec{p}_{\pi^-}^\prime-\vec{p}_{K^+}^\prime = 2\vec{k}+\vec{q}$ form used in Refs.\cite{Pavao:2017kcr} for the VPP vertex instead of $P-2\vec{k}-\vec{q}$ in  Ref.\cite{Aceti:2016yeb}. }
\label{tabcij}
\end{table}
%where the the $ t^{(b)(c)}$ is similar to the Eq.\ref{Eq:b8}, the difference are the extra propagator $a_1$ and a coupling of $g_{ \bar{K}^{*}K a_1}$.
We have $g_{ \bar{K}K f_0}= g_{ K^+K^- f_0}=-g_{ \bar{K}^0K^0 f_0}$ for the coupling between $\bar{K}K $ and $f_0$ in isospin base.
The value  of coupling $g_{ \bar{K}^{*}K a_1}$ in isospin base has been give in Ref. \cite{Roca:2005nm}
\begin{equation}
g_{ \bar{K}^{*}K a_1} =  (1872 - i 1486)\ \text{MeV}.
\end{equation}
Meanwhile, the coupling between $a_1^+$ and the combination with $ I=1, C=+$ and $G=- $ of the $\bar{K}^* K$ pair is represented by the state
\begin{equation}
\frac{1}{\sqrt{2}}(\bar{K}^* K - K^* \bar{K})  =   \frac{1}{\sqrt{2}}(\bar{K}^{*0} K^+ - K^{*+} \bar{K}^0).
\end{equation}
The $C^{\prime}$ in the $t^{(b)(c)}$ stand for the effective coupling strength of  the $D^{0} \to  a_1^{+} \pi^-$ vertice. In the derivation of $C^{\prime}$, combining Br$(D^0 \to a_1^+(1260), a_1 \to 2\pi^+ \pi^-) = (4.52 \pm 0.31) \times 10^{-3}$ from PDG and  Refs. \cite{FOCUS:2007ern,Wang:2015cis}, we take the branch ratio as Br$(D^0 \to a_1^+ \pi^-)= \left(4.1 \pm 0.4 \right)\times 10^{-3}$.
For the propagator of $a_1$ in Eq.(\ref{M}), we take parameters of $a_1$ as $m_{a_1}=1230$ MeV and $\Gamma_{a_1}=425$ MeV, coming from the use of Ref.\cite{Xie:2016lvs}.

Meanwhile, we further consider that the $f_0(980)$ decays to the $\pi^+ \pi^-$ final state, as depicted in Fig.\ref{fig1}(d).
In order to write the amplitude of $D^0 \to \pi^+ \pi^- f_0(980), f_0 \to \pi^+ \pi^- $ process, following Refs.\cite{Liang:2019jtr,Sakai:2017hpg}, we only need to replace the couplings $C_3^a g_{ \bar{K}K f_0}$ or $g_{K^+ K^- f_0 }$ in Eqs.(\ref{M},\ref{Gamma-1}) by the transition amplitude $t_{K^+ K^-\to \pi^+ \pi^-}$.
The amplitude $t_{K^+ K^-\to \pi^+ \pi^-}$ is obtained by solving the coupled channels Bethe-Salpeter (BS) equation in the chiral unitary approach, in which $f_0$ appears as dynamically generated state.
The BS equation is given by
\begin{equation}\label{Eq:b20}
T=\left[1-VG \right]^{-1}V,
\end{equation}
where the $V$ and $G$ are the  interaction potential and meson loop function respectively, which have been calculated in Ref.\cite{Liang:2014tia}.
The meson loop function $G$ is regulated by a cut-off $q_{\text{max}}=600$ MeV.
Finally, the double differential distribution for $D^0 \to \pi^+ \pi^- f_0(980), f_0 \to \pi^+ \pi^- $ decay process is written as
\begin{equation}\label{Eq:b21}
\begin{aligned}
\frac{d^2 \Gamma^\prime}{d M_{inv}(\pi f_0) d M_{inv}(\pi^+ \pi^-)}&=\frac{g^2 C^2}{(2\pi)^5}  \frac{ |\vec{p}_{\pi^+}||\vec{\tilde{q}}_{\pi}||\vec{k}|}{4m_{D^0}^2}|t_T|^2\cdot |t_{K^+ K^-\to \pi^+ \pi^-}|^2. \\
\end{aligned}
\end{equation}

Similar to Eq.\ref{M}, the amplitudes expressions of Figs.\ref{fig1}(d)(e)(f) have
\begin{equation}\label{Mdef}
\begin{aligned}
t^{d} & =  -  C (C_2^a\ g)  \ \vec{p}_{\pi^+}^\prime  \cdot \vec{k} \times  \tilde{t}_T(m_{K^{*0}},m_{K^-}, m_{K^+}) \ (t_{K^+ K^-\to \pi^+ \pi^-}), \\
t^{e} & = - C^\prime \frac{1}{M_{inv}^2(\pi f_0)-m_{a_1}^2+im_{a_1}\Gamma_{a_1}} (C_1^b \ g_{ \bar{K}^{*}K a_1} ) (C_2^b\ g) \  \vec{p}_{\pi^-}^\prime  \cdot \vec{k} \times  \tilde{t}_T(m_{\bar{K}^{*0}},m_{K^+}, m_{K^-})\  (t_{K^+ K^-\to \pi^+ \pi^-}), \\
t^{f} & = - C^\prime \frac{1}{M_{inv}^2(\pi f_0)-m_{a_1}^2+im_{a_1}\Gamma_{a_1}} (C_1^c\ g_{ \bar{K}^{*}K a_1} ) (C_2^c\ g) \  \vec{p}_{\pi^-}^\prime  \cdot \vec{k} \times  \tilde{t}_T(m_{K^{*+}},m_{\bar{K}^0}, m_{K^0})\  (t_{\bar{K}^0 K^0\to \pi^+ \pi^-}). \\
\end{aligned}
\end{equation}

\section{RESULTS}
\label{III}

\begin{figure}[h]
\subfigure[]{
\includegraphics[scale=0.3]{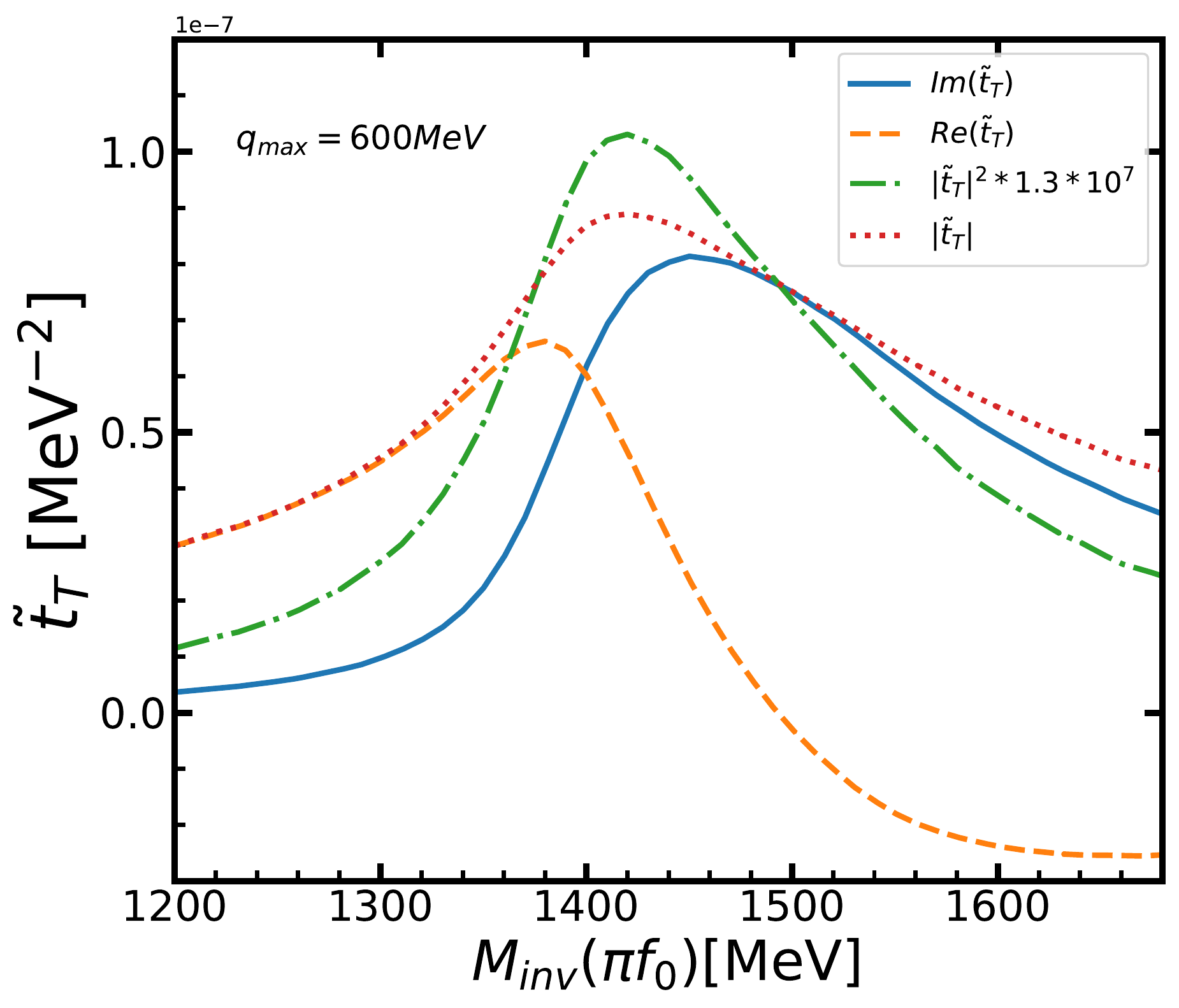}
}
\subfigure[]{
\includegraphics[scale=0.3]{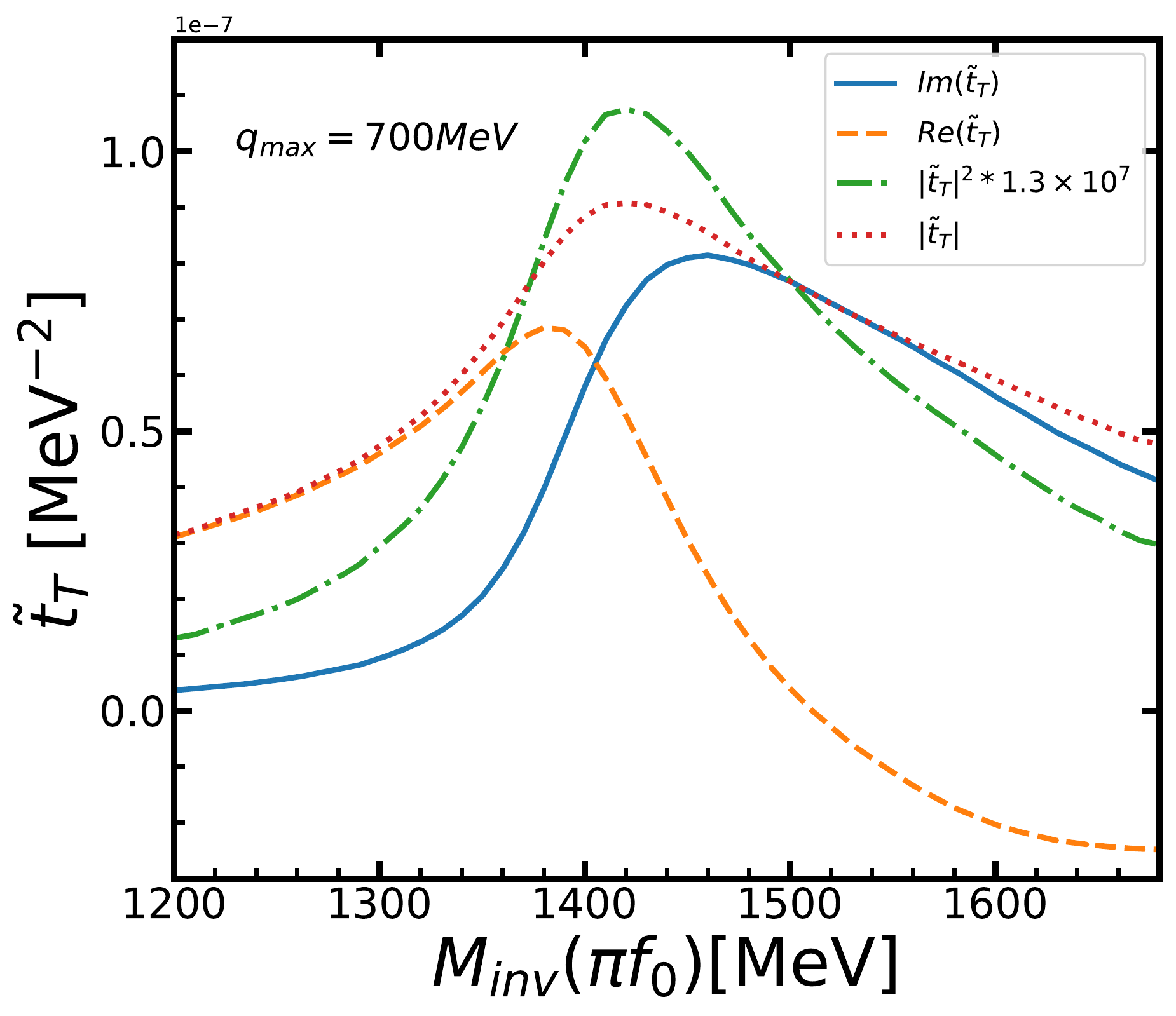}
}
\subfigure[]{
\includegraphics[scale=0.3]{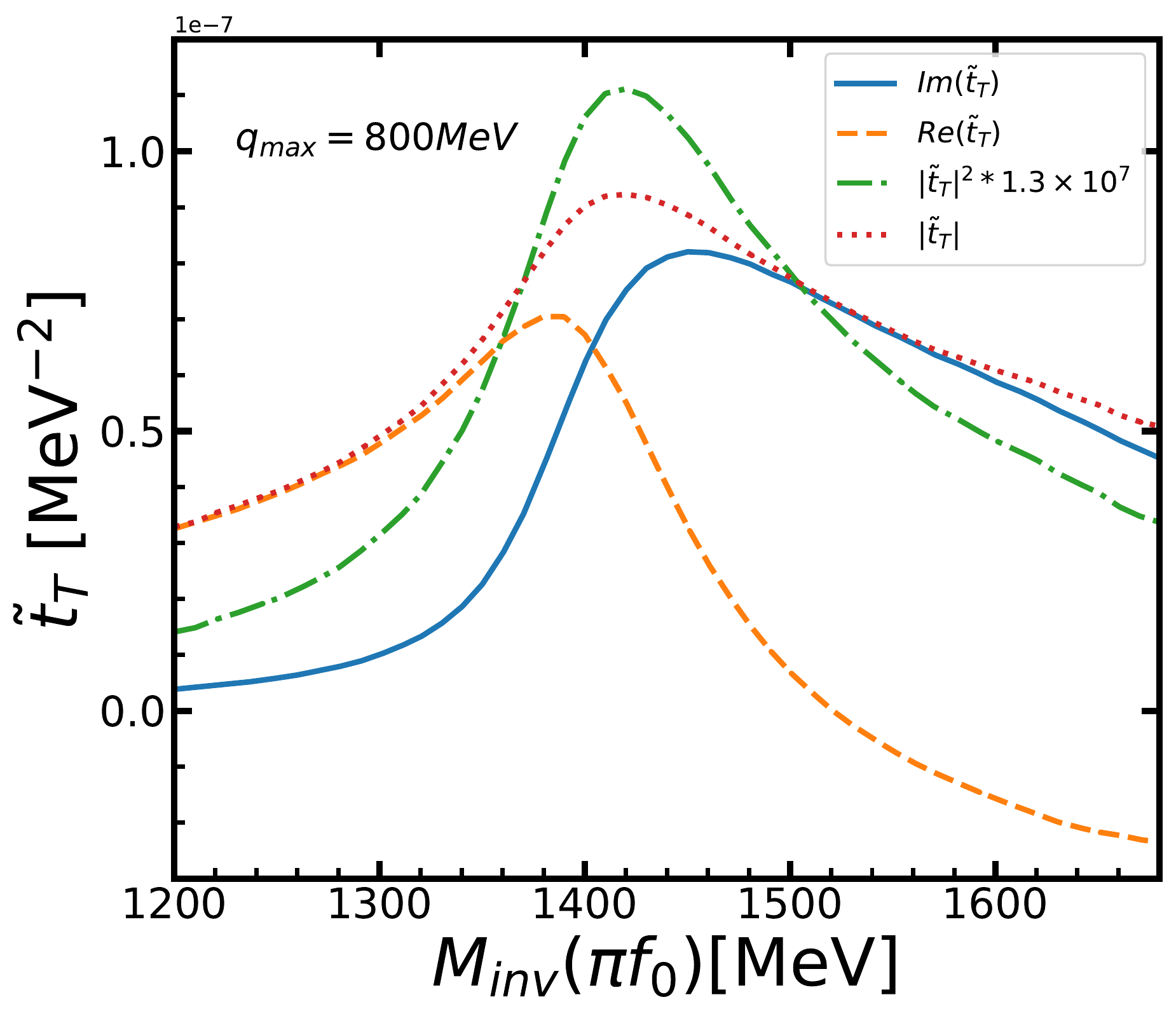}
}
\caption{
The distributions of the triangle amplitude $\tilde{t}_T$ for the $K^{*0} K^+ K^-$ triangle loop in the Fig.\ref{fig1}(a). We take $M_{f_0}=980$ MeV, and times $1.3\cdot 10^{7}$ for $t_T^2$.}
\label{Fig-tT}
\end{figure}

With the former formula Eq.(\ref{Eq:b11}), by fixing the value of $M_{f_0}$, the distributions of the triangle amplitudes $\tilde{t}_T$, Im$(\tilde{t}_T)$, Re$(\tilde{t}_T)$ and $|\tilde{t}_T|^2 \times 1.3\cdot10^{7}$ as functions of the $M_{inv}(\pi f_0)$ invariant mass have been shown in Fig.\ref{Fig-tT} for the $K^{*0} K^+ K^-$ triangle loop diagram.
From the Fig.\ref{Fig-tT}, we see that there is a peak around $1418$ MeV for the $|\tilde{t}_T|^2$
%with width of about $200$ MeV
, which is consistent with the result of Eq.(\ref{Eq:2-1}).
Since the kinematic factors are functions of the invariant mass $M_{inv}(\pi f_0)$, as $M_{inv}(\pi f_0)$ increases, it could impose restrictions on phase space and change the shape of the final mass distribution.
According to Ref.\cite{Pavao:2017kcr}, we need to consider whether the cutoff $q_{max}$ and triangle singularity will affect the behavior of Im$(\tilde{t}_T)$.
Comparing the three sub-figures of Fig.\ref{Fig-tT} where $M_{inv}(\pi f_0)$ is fixed at 600, 700 and 800 MeV respectively, one can find that as the $q_{max}$ increases the behavior of the Im$(\tilde{t}_T)$ at the higher $M_{inv}(\pi f_0)$ become softer, while the peak that was associated to the triangle singularity remains.
There are two peaks for Im$(\tilde{t}_T)$ and Re$(\tilde{t}_T)$ located at $1440$ MeV and $1390$ MeV, respectively. The reason is that the peak of Im$(\tilde{t}_T)$ is derived from the triangle singularity while the peak of Re$(\tilde{t}_T)$ is derived from the threshold of the $K^{*0} K^\pm$ that is about $1390$ MeV in Fig.\ref{fig1}(a).
Next, we will more intuitively show the behavior of Im$(\tilde{t}_T)$ and Re$(\tilde{t}_T)$ peaks.
%The former is consistent with the position of triangle singularity, and the latter comes from the $K^{*0} K^\pm$ threshold.

\begin{figure}[h]
\subfigure[]{
\includegraphics[scale=0.35]{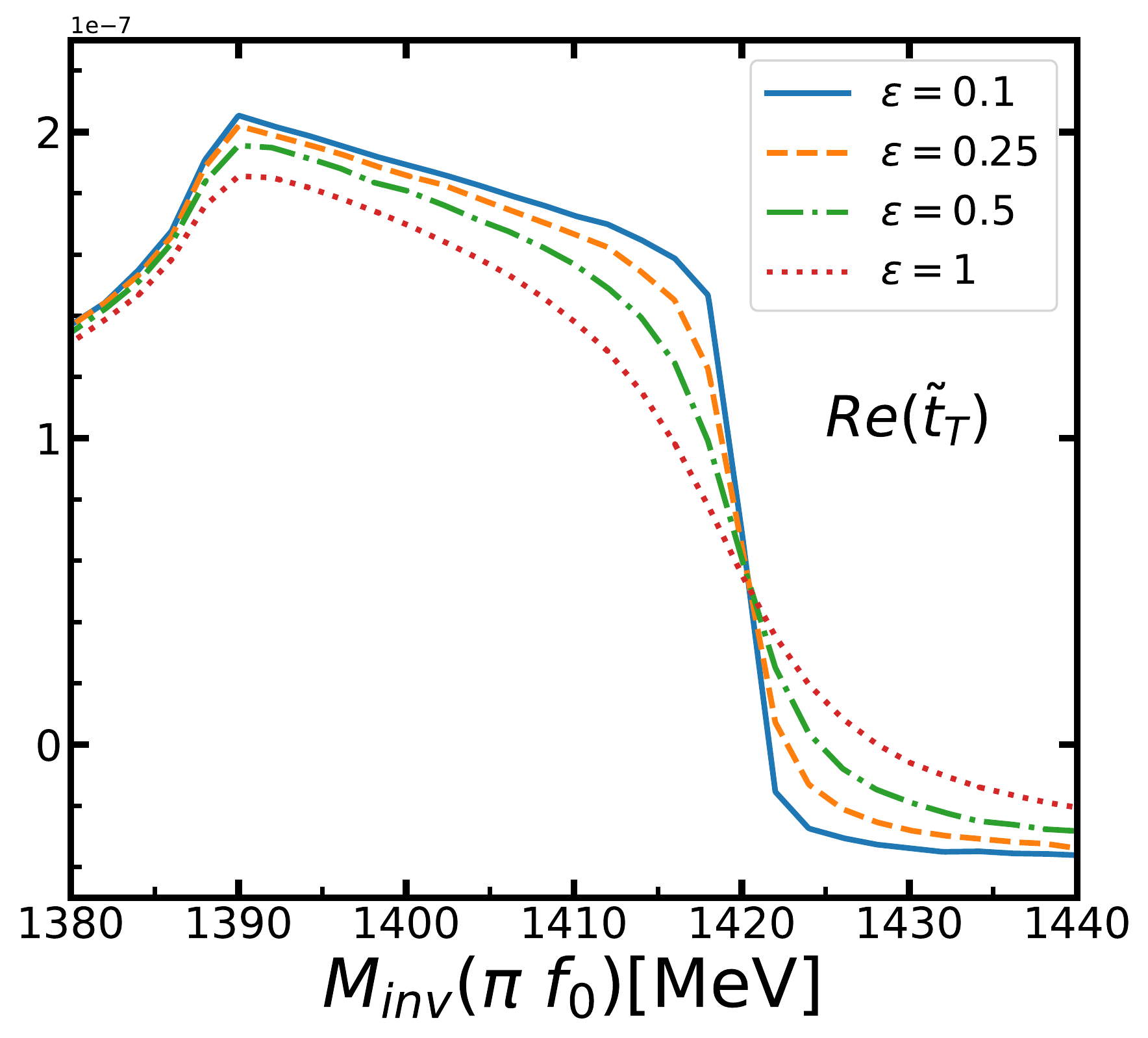}
}\qquad
\subfigure[]{
\includegraphics[scale=0.35]{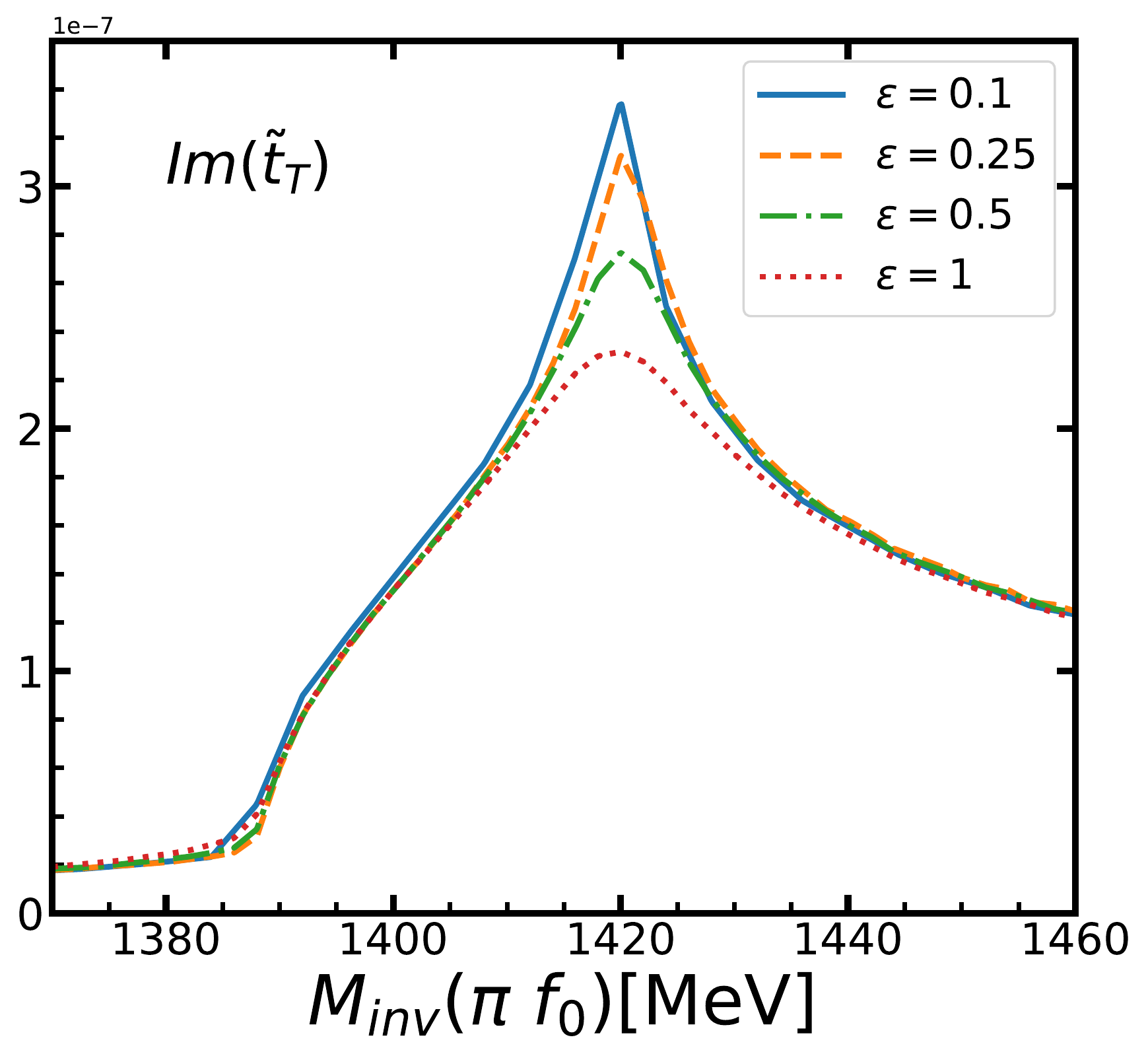}
}
\caption{
The distributions of the real (left) and imaginary (right) parts of $\tilde{t}_T$ for the $K^{*0} K^+ K^-$ triangle loop by taking $M_{f_0}=990$ MeV and the different values of $\Gamma_{K^*}/2=\epsilon=1,\ 0.5,\ 0.25$ and $0.1$ MeV.}
\label{Re-Im}
\end{figure}

In Fig.\ref{Re-Im}, by analogy to Ref.\cite{Pavao:2017kcr}, we show the development of the triangle singularity of $K^{*0} K^+ K^-$ triangle loop with $\Gamma_{K^*}/2$ and $\epsilon$ fixed at different finite values but close to zero.
To reach the triangle singularity in the Fig.\ref{fig1}(a), it is required that the mass of $f_0$ should be slightly larger than the $K^- K^+$ threshold as in Eq.(\ref{condition}).
In Fig.\ref{Re-Im}, we have taken $m_{f_0}=990$ MeV.
Due to the different sources of the two peaks of Im$(\tilde{t}_T)$ and Re$(\tilde{t}_T)$, from  Fig.\ref{Re-Im}, we can see that there are two different distribution behavior for Re$(\tilde{t}_T)$ and Im$(\tilde{t}_T)$.
For the Re$(\tilde{t}_T)$, in Fig.\ref{Re-Im}(a), there is a cusp located at threshold of the $K^{*0}K$ system $1390$ MeV and a sharp downfall nearby the triangle singularity 1418 MeV.
And, in Fig.\ref{Re-Im}(b), the triangle singularity  appeares in the form of a narrow peak around 1418 MeV in the distribution of Im$(\tilde{t}_T)$.
At the same time, by taking the different values of $\Gamma_{K^*}/2$ and $\epsilon$, namely 1, 0.5, 0.25 and 0.1 MeV, Fig.\ref{Re-Im} shows that as the values decrease, the cusp of the Re$(\tilde{t}_T)$ converges to a finite value, which is associated to the threshold of the $K^{*0}K$, and the peak of Im$(\tilde{t}_T)$ becomes more and more sharp and finally transforms into a singularity when $\Gamma_{K^*}/2=\epsilon=0$.

\begin{figure}[h]
\subfigure[]{
\includegraphics[scale=0.29]{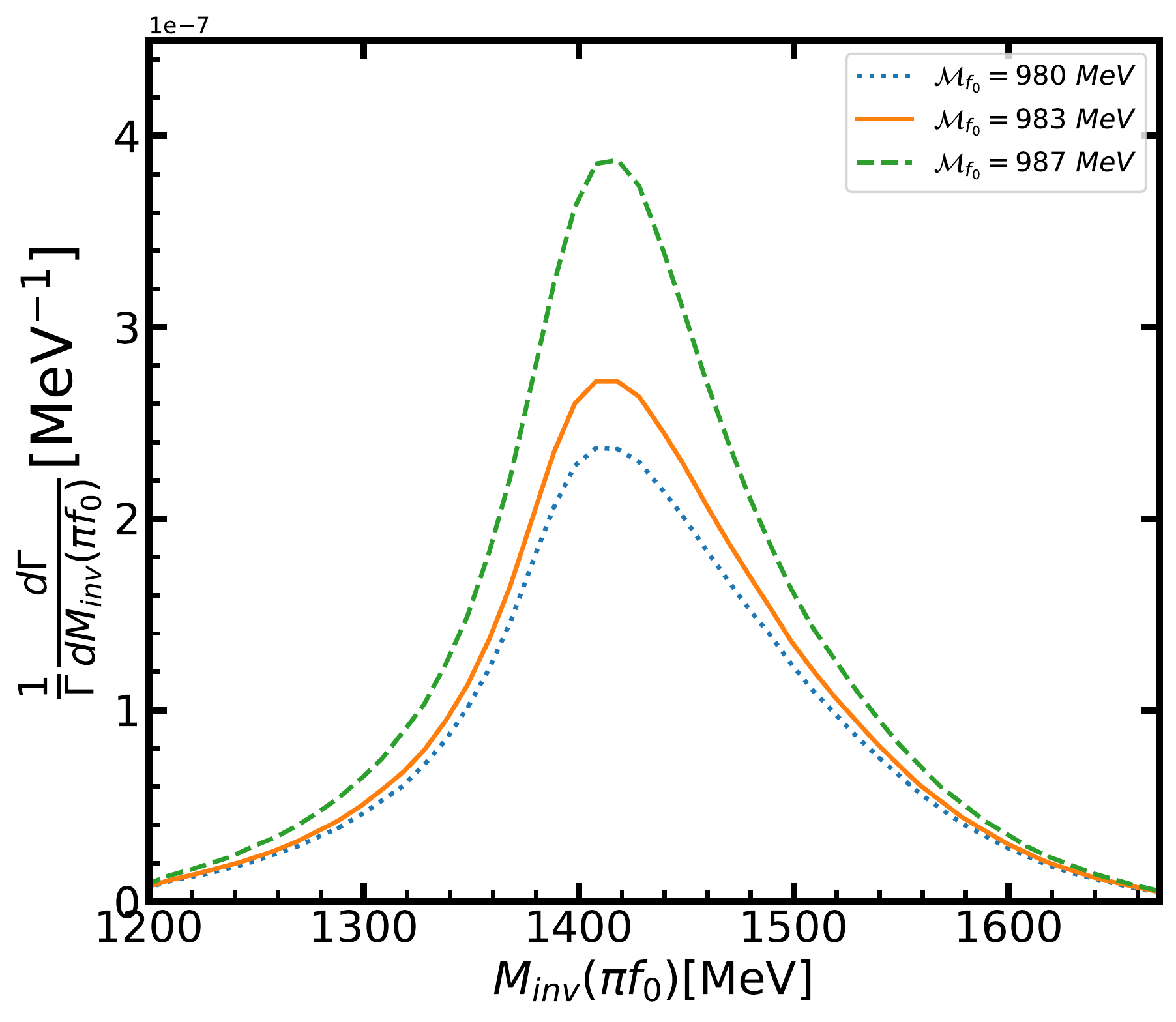}
}\qquad
\subfigure[]{
\includegraphics[scale=0.29]{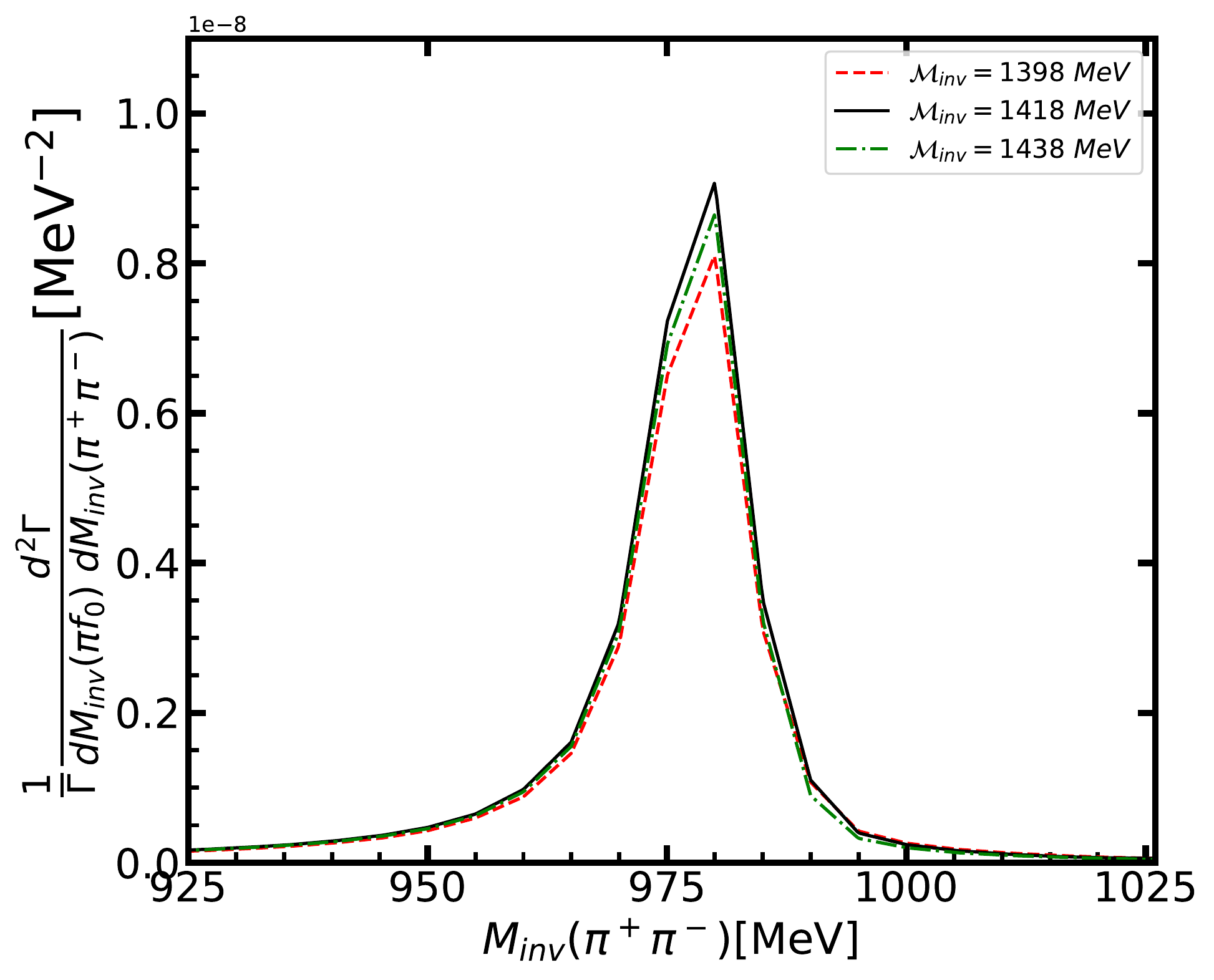}
}
\subfigure[]{
\includegraphics[scale=0.29]{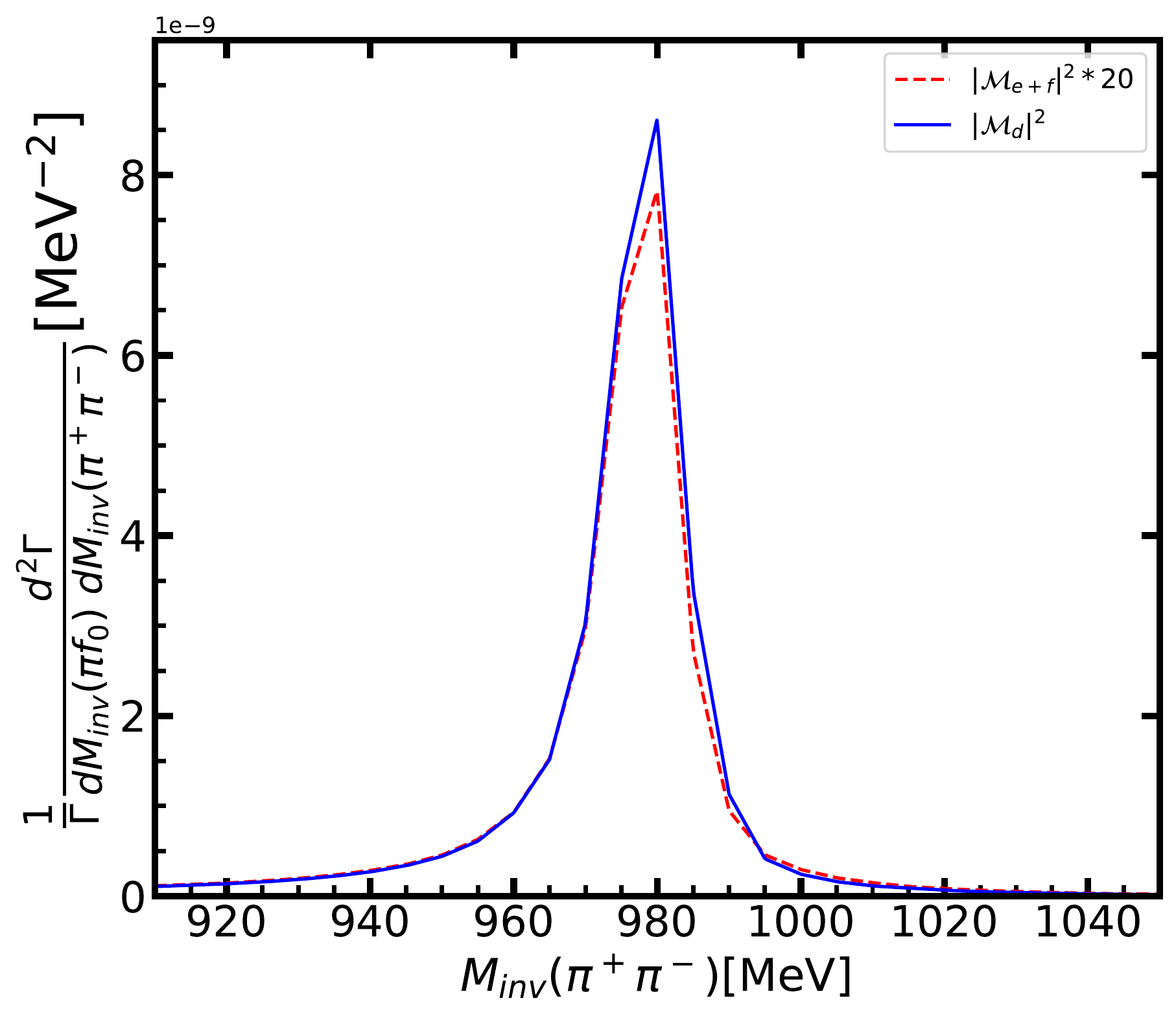}
}
\caption{
a): The differential distribution of decay width $\Gamma^\prime$  for $f_0(980)$ productions, as Eq.\ref{Gamma-1}.
b): The double differential distribution of decay width $\Gamma$ as a function of the invariant mass $M_{inv}(\pi^+ \pi^-)$ for the four body decay processes at $M_{inv}(\pi f_0)=1398$, 1418 and 1438 MeV.
c): The differential distributions of $\Gamma$ for Fig.\ref{fig1}(d) and  Fig.\ref{fig1}(e) plus Fig.\ref{fig1}(f) with $M_{inv}(\pi f_0)=1418$ MeV.}
\label{Mpipi}
\end{figure}
\begin{figure}[h]
\subfigure[]{
\includegraphics[scale=0.35]{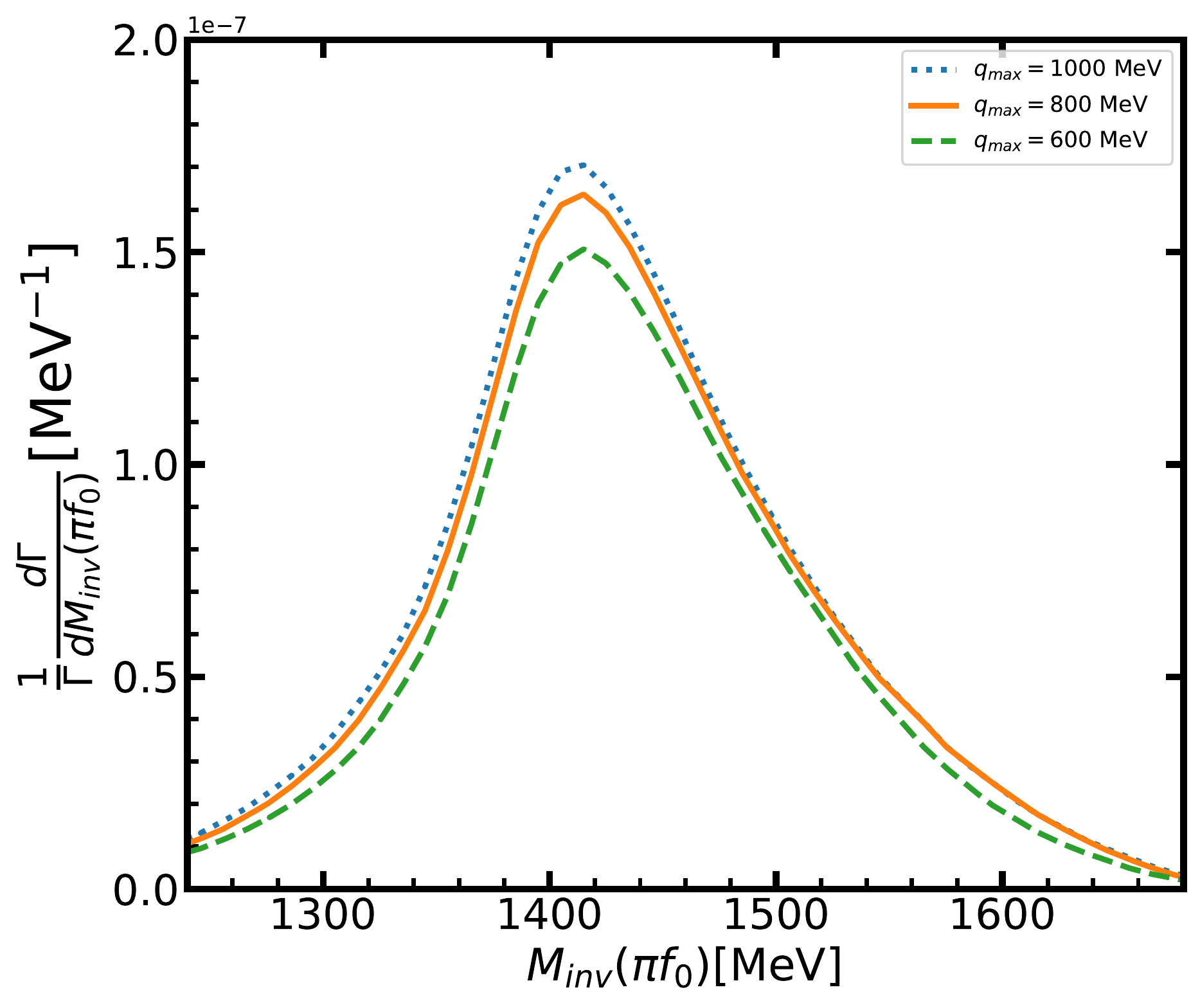}
}
\subfigure[]{
\includegraphics[scale=0.35]{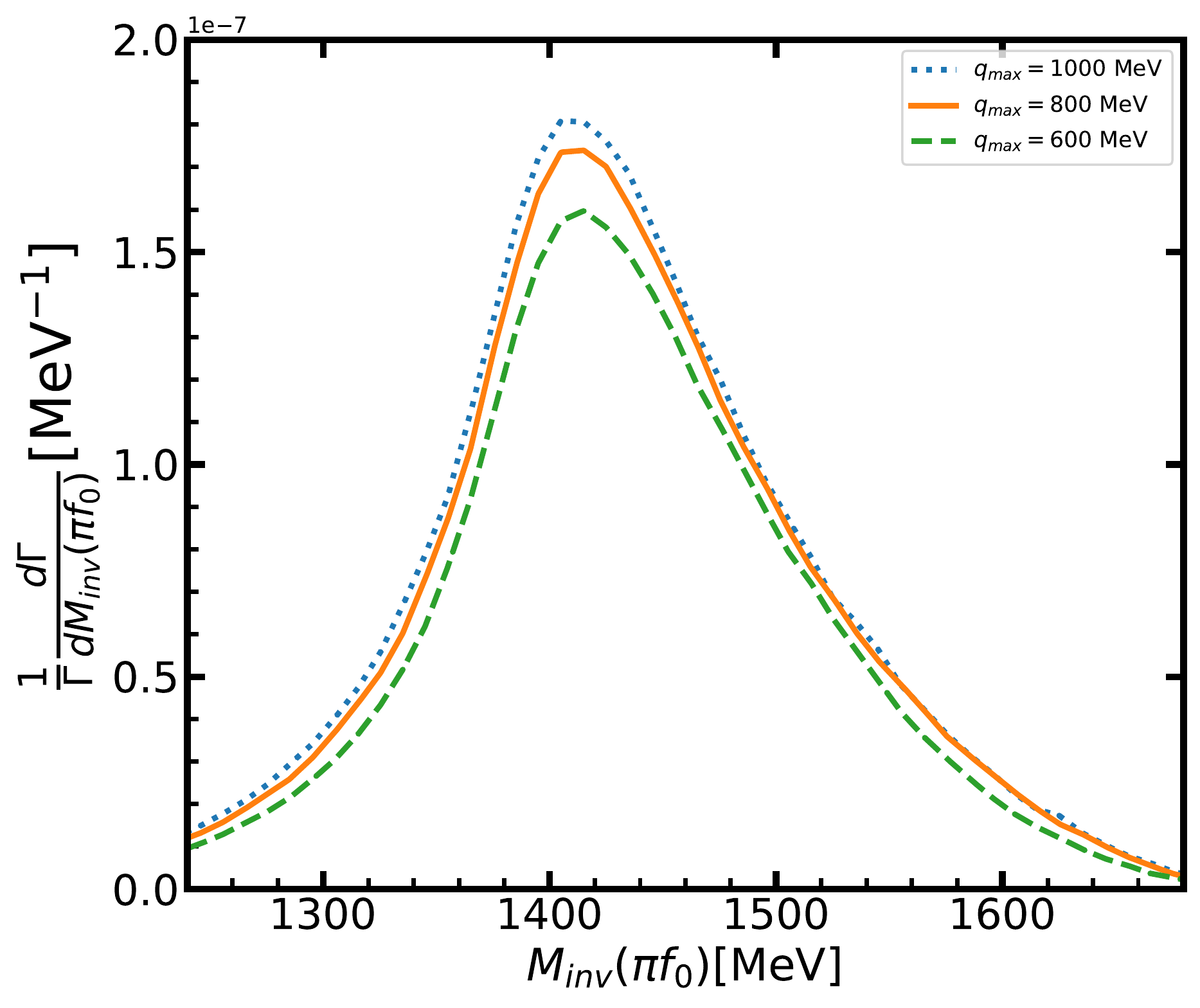}
}
%\subfigure[]{
%\includegraphics[scale=0.35]{fig6b.pdf}
%}
\caption{
The differential distribution of decay width $\Gamma$ as a function of the invariant mass $M_{\pi f_0}$ for the four body decay processes.
a) The integration region  is $M_{f_0}\in [900,1000]$ MeV.
b) The integration region  is $M_{f_0}\in [800,1200]$ MeV.
%a) The contributions of total and  Fig.\ref{fig1}(b) amplitudes. b) The contributions of Figs.\ref{fig1}(d)(e) amplitudes.
}
\label{Gamma}
\end{figure}

As shown in Fig.\ref{Mpipi}(a), differential distributions of decay width $\Gamma^\prime$ have been depicted for different $f_0(980)$ masses $M_{f_0} =$ 980, 983 and 987 MeV.
It is clear that the mass of $f_0(980)$ will enhance the result of decay width, but the peaks of differential distributions are still located around the location of the triangle singularity $M_{inv}(\pi f_0)=1418$ MeV.
In Fig.\ref{Mpipi}(b), we plot the double differential distribution of decay width $ \frac{1}{\Gamma_{D^0}}\frac{d^2 \Gamma}{d M_{inv}(\pi f_0)d M_{inv}(\pi^- \pi^+)}$ as a function of the invariant mass $M_{inv}(\pi^- \pi^+)$ in the region of the $f_0(980)$ for the four body decay process.
In Fig.\ref{Mpipi}(b), we take  $M_{inv}(\pi f_0)=1398$, 1418 and 1438 MeV as the  triangle singularity is around.
%Compared with Figs.\ref{Mpipi}(a)(b), it is obvious that the contributions to width $\Gamma$ mainly come from  the $K^* KK$ triangle loop diagrams.
For the Fig.\ref{Mpipi}(b), there is a clear peak around $980$ MeV, and have a strong contribution to the $M_{inv}(\pi^+ \pi^-)$ distribution around the region of $M_{inv}(\pi^+ \pi^-)=980$ MeV.
The Fig.\ref{Mpipi}(c) shows that although we consider the contributions of the intermediate state $a_1$, it's very small compared with the Fig.\ref{fig1}(a).
Moreover, the peak at the triangle singularity ($M_{inv}(\pi f_0)=1418$) is significantly larger than the values of $M_{inv}(\pi f_0)=1398$ and 1438 MeV.
In the following analysis, we only focus on the domain of $f_0(980)$, and the main contribution comes from the range of $M_{f_0}\in [900,1000]$.
Therefore, we can restrict the integral range of $M_{inv}(\pi^+ \pi^-)$ of Eq.(\ref{Eq:b21}) to these limits.
Further, we can obtain the differential distribution for $\frac{1}{\Gamma_{D^0}}\frac{d \Gamma}{d M_{inv}(\pi f_0)}$ as a function of the invariant mass $M_{inv}(\pi f_0)$.  The results have been depicted in Fig.\ref{Gamma}. From the Fig.\ref{Gamma}, for the total contribution, there is a clear peak located at 1418 MeV  considering the $f_0(980)$ as a dynamically generated states, which is consistent with Figs.(\ref{Fig-tT})(\ref{Mpipi}).
%It is obvious that the contributions of decay width are dominated by the $K^*KK$ triangle loop diagram, while the contribution of the $\sigma \pi \pi$ and $\rho \pi \pi$ loop is about two orders of magnitude smaller than that of the $K^*KK$ triangle loop.

\begin{comment}
Finally, from the PDG \cite{Zyla:2020zbs,Link_2007}, we can get the experiment result for $D^0 \to f_0(980) \pi^+ \pi^-,\ f_0 \to \pi^+ \pi^- $ decay process
\begin{equation}
\label{result-exp}
\text{Br}_{exp}(D^0 \to \pi^+ \pi^- f_0,f_0\to \pi^+ \pi^-)= \left(1.8 \pm 0.5 \right) \times 10^{-4}.
\end{equation}
In order to compare with this data, we can integrate Fig.\ref{Gamma} over invariant mass $M_{inv}(\pi f_0)$, the theoretical decay branching ratio for $D^0 \to \pi^+ \pi^- \pi^+ \pi^-$ process is obtained as
\begin{equation}\label{result-theo}
\text{Br}_{theory}(D^0 \to \pi^+ \pi^- f_0,f_0\to \pi^+ \pi^-)= 1.446  \times 10^{-4}.
\end{equation}
Form the result of Eq.(\ref{result-theo}), we get a result that can be compared with experiments Eq.(\ref{result-exp}).
We need to note that the integration range of $M_{inv}(\pi^+ \pi^-)$ is selected the range of $[900,1000]$ MeV only.
Meantime, in the experiments, some background contributions must be considered.
\end{comment}

\section{SUMMARY}
\label{IV}

In the present work, we give the derivation details and formalism for the decay width calculation of $D^0 \to \pi^+ \pi^- f_0(980),\ f_0 \to \pi^+ \pi^-$ process. From the Figs.\ref{fig1}(a), $D^0$ first decays to $K^{*0} \pi K$, then the $K^{*0}$ decays to the $\pi \bar{K}$ and the $\bar{K} K$ fuse to form a $f_0(980)$. Only the triangle mechanism corresponding to $K^{*0} K\bar{K}$ triangle diagrams can produce  triangle singularity around $1418$ MeV for the invariant mass of $M_{inv}(\pi f_0)$. Through calculating the $\frac{d^2 \Gamma}{d M_{inv}(\pi f_0) d M_{inv}(\pi^+ \pi^-)}$, a clear peak is produced in the invariant mass distribution of $\pi^+ \pi^-$ system that comes from the $f_0$ decay, showing a clear $f_0(980)$ shape.
After integrating over $M_{inv}(\pi^+ \pi^-)$, the  $\frac{d \Gamma}{d M_{inv}(\pi f_0)}$ shows a clear peak in the $ M_{inv}(\pi f_0)$ invariant mass distribution located at 1418 MeV, and the main contribution of the peak comes from the triangle singularity.
%Finally, we can obtain a theoretical result of decay branching $\text{Br}(D^0 \to \pi^+ \pi^- f_0,f_0\to \pi^+ \pi^-)= 1.446  \times 10^{-4}$  which is possible to be compared with the experimental result.

\begin{acknowledgments}
The authors thank professor Weihong Liang and professor Eulogio Oset for their patient guidance. Hao Sun is supported by the National Natural Science Foundation of China (Grant No.12075043, No.12147205).
\end{acknowledgments}

\bibliography{ref}

\end{document}